\documentclass[aps,prc,preprint,showpacs,superscriptaddress]{revtex4}  
\usepackage{graphicx}
\usepackage{amsmath}
\begin{document}
\newcommand {\nc} {\newcommand}
\nc {\beq} {\begin{eqnarray}} 
\nc {\eol} {\nonumber \\} 
\nc {\eeq} {\end{eqnarray}} 
\nc {\eeqn} [1] {\label{#1} \end{eqnarray}}   
\nc {\eoln} [1] {\label{#1} \\} 
\nc {\ve} [1] {\mbox{\boldmath $#1$}}
\nc {\rref} [1] {(\ref{#1})} 
\nc {\Eq} [1] {Eq.~(\ref{#1})} 
\nc {\la} {\mbox{$\langle$}}
\nc {\ra} {\mbox{$\rangle$}}
\nc {\dd} {\mbox{${\rm d}$}}
\nc {\cM} {\mathcal{M}} 
\nc {\cY} {\mathcal{Y}} 
\nc {\dem} {\mbox{$\frac{1}{2}$}}
\nc {\ut} {\mbox{$\frac{1}{3}$}} 
\nc {\qt} {\mbox{$\frac{4}{3}$}} 
\nc {\Li} {\mbox{$^6\mathrm{Li}$}}
\nc {\M} {\mbox{$\mathcal{M}$}}
\title{Astrophysical $S$-factor of the direct $\alpha(d,\gamma)^6$Li capture reaction in a three-body model}
%
%
\author {E. M. Tursunov}
\email{tursune@inp.uz} \affiliation {Institute of Nuclear Physics,
Academy of Sciences, 100214, Ulugbek, Tashkent, Uzbekistan}
\author{D. Baye}
\email{dbaye@ulb.ac.be} \affiliation {Physique Quantique, and Physique Nucl\'eaire Th\'eorique et Physique Math\'ematique, \\ 
C.P. 229, Universit\'e libre de Bruxelles (ULB), B-1050 Brussels Belgium}
\author {S.A. Turakulov}
\email{turakulov@inp.uz} \affiliation {Institute of Nuclear Physics,
Academy of Sciences, 100214, Ulugbek, Tashkent, Uzbekistan}
%
\begin{abstract}
At the long-wavelength approximation, electric dipole transitions are forbidden between isospin-zero states. 
In an $\alpha+n+p$ model with $T = 1$ contributions, 
the $\alpha(d,\gamma)^6$Li astrophysical $S$-factor is in agreement 
with the experimental data of the LUNA collaboration, without adjustable parameter. 
The exact-masses prescription used to avoid the disappearance 
of $E1$ transitions in potential models is not founded at the microscopic level. 
\end{abstract}
\maketitle
\section{Introduction}
A radiative-capture reaction is an electromagnetic transition between an initial scattering state 
and a final bound state. 
Astrophysical collision energies can be very low with respect to the Coulomb barrier 
and cross sections are then tiny. 
The dominant multipolarity is $E1$ in general.
In the special case of reactions between $N = Z$ nuclei however, 
$E1$ transitions are « forbidden » by an isospin selection rule at the long-wavelength approximation (LWA) 
and $E2$ transitions become crucial. 
Nevertheless, $E1$ transitions are not exactly forbidden since isospin is an approximate symmetry. 
The analysis of the recent LUNA data \cite{ATM14,TAA17} for the $\alpha(d,\gamma)^6$Li reaction 
indicates that $E1$ cross sections dominate the $E2$ cross sections below about 0.1 MeV.  
Since $E1$ transitions vanish in many models, 
recent calculations use the « exact-masses » prescription to avoid their disappearance \cite{TKT16}. 
Here we present results of the simplest model allowing $E1$ transitions 
thanks to small $T = 1$ components, i.e.\ the $\alpha + n + p$ three-body model. 
\section{Isospin-forbidden $E1$ transitions}
The electric multipole operators read at the LWA, 
\beq
{\cal M}_\mu^{E\lambda} = e \sum_{j=1}^A (\dem - t_{j3}) r^{\prime\lambda}_j Y_{\lambda\mu} (\Omega'_j),
\eeqn{1}
where $A$ is the number of nucleons, $\ve{r}'_j = (r'_j,\Omega'_j)$ is the coordinate of nucleon $j$ 
with respect to the centre of mass of the system, 
and $t_{j3}$ is the third component of its isospin operator $\ve{t}_j$. 
The isoscalar (IS) part of the $E1$ operator vanishes at the LWA 
since $\sum_{j=1}^A r'_j Y_{1\mu} (\Omega'_j) = 0$ and 
this operator becomes an isovector (IV), 
\beq
{\cal M}_\mu^{E1} = -e \sum_{j=1}^A t_{j3}\, r'_j Y_{1\mu} (\Omega'_j) = {\cal M}_\mu^{E1,{\rm IV}}.
\eeqn{2}
At the LWA, $E1$ matrix elements thus vanish between isospin-zero states. 
This leads to the total isospin $T$ selection rule in $N = Z$ nuclei and reactions: 
$T_i = 0 \rightarrow T_f = 0$ is forbidden. 
But $E1$ transitions are not exactly forbidden in these $N = Z$ systems 
because isospin is not an exact quantum number. 
Small $T = 1$ admixtures appear in the wave functions.   
The main isovector $E1$ contributions are due to $T_f = 1$ admixtures in the final state 
or to $T_i = 1$ admixtures in the initial state. 
Moreover, the isoscalar $E1$ operator reads beyond the LWA,  
\beq
{\cal M}_\mu^{E1,{\rm IS}} & \approx & 
- \frac{1}{60} e k_\gamma^2 \sum_{j=1}^A r_j^{\prime 3} Y_{1\mu} (\Omega'_j),
\eeqn{3}
up to terms that should give only a small contribution \cite{Ba12}, 
contrary to other expressions often used in the literature. 
The isoscalar $E1$ contribution to the capture involves the $T = 0$ parts of the wave functions.
\section{Three-body model of $\alpha(d,\gamma)^6$Li reaction}
The present wave functions \cite{BT18} are adapted from the $\alpha + n + p$ model of Ref.~\cite{TKT16}. 
The $J_f = 1^+$ final bound state is described in hyperspherical coordinates 
and the initial scattering states are described in Jacobi coordinates. 
Three-body effective $E1$ and $E2$ operators are constructed 
which assume that the $\alpha$ particle or cluster is in its $0^+$ ground state. 
For example, the isovector part of the effective three-body $E1$ operator reads at the LWA, 
\beq
\widetilde{\cM}_\mu^{E1,{\rm IV}} = \dem e r Y_{1\mu}(\Omega_r), 
\eeqn{4}
where $\ve{r}$ is the Jacobi coordinate between $n$ and $p$. 
The expressions of the isoscalar part of the $E1$ operator beyond the LWA and of the $E2$ operator 
can be found in Ref.~\cite{BT18}.  

The three-body states contain $S = 0$ and 1 components. 
Because of the isospin zero of the $\alpha$ particle 
and the antisymmetry of the $n + p$ subsystem with orbital momentum $l$, 
the components with $l + S$ odd correspond to $T = 0$ and those with $l + S$ even to $T = 1$. 
The initial scattering state is described by the product of a frozen deuteron wave function 
($l_i = 0$, $S_i = 1$) and $\alpha + d$ $L$ partial scattering waves. 
Hence, it is purely $T_i = 0$.  
The $J_f = 1^+$ final bound state contains a small $T_f = 1$ component (about 0.5 \%). 
The $E1$ transitions start from $L_i = 1$ and the $E2$ transitions from $L_i = 0$ and 2. 

\begin{figure}[htb]
\setlength{\unitlength}{1 mm}
\begin{picture}(160,50) (-10,10) 
\put(0,0){\includegraphics[width=0.8\textwidth]{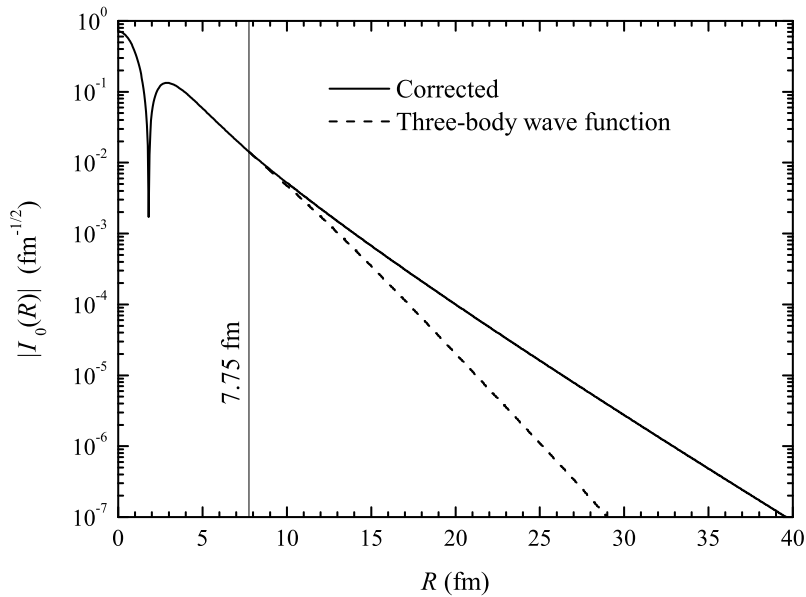}}
\end{picture} \\
\caption{$E2$ overlap integral $|I_0(R)|$ with and without correction beyond 7.75 fm.}
\label{fig:1}
\end{figure}
This model requires an asymptotic correction to the $E2$ matrix elements. 
Indeed the overlap integrals $I_L(R)$ of the deuteron and $\alpha + n+ p$ final wave functions 
decrease too fast beyond 10 fm as shown for $L = 0$ by Fig.~\ref{fig:1}. 
This is corrected by matching at 7.75 fm the overlap integrals with the exact Whittaker asymptotic function 
multiplied by realistic asymptotic normalization coefficients. 

\begin{figure}[thb]
\setlength{\unitlength}{1 mm}
\begin{picture}(160,50) (-10,10) 
\put(0,0){\includegraphics[width=0.8\textwidth]{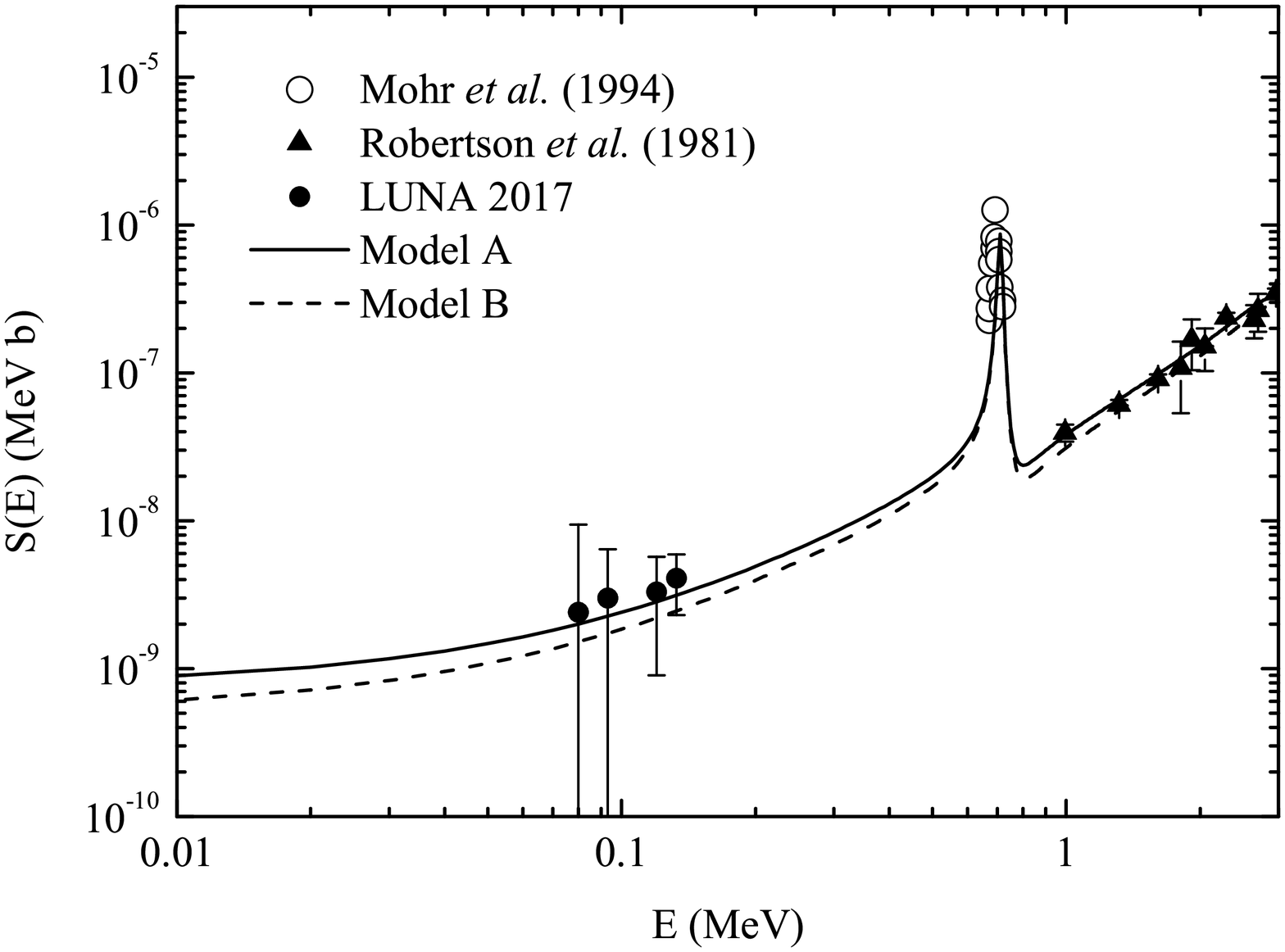}}
\end{picture} \\
\caption{Total $E1+E2$ astrophysical $S$ factor (full and dashed lines). 
Experimental data from Refs.~\cite{robe81} (triangles), 
\cite{mohr94} (open circles), and \cite{TAA17} (full circles). 
Adapted from Ref.~\cite{BT18}. }
\label{fig:2}
\end{figure}

Total $E1+E2$ astrophysical $S$ factors calculated in the three-body model with the $E2$ correction 
are compared in Fig.~\ref{fig:2} with experimental data. 
The isoscalar $E1$ capture contribution is small and can be neglected in first approximation. 
The isovector $E1$ contribution dominates below about 0.1 MeV. 
Model A and B correspond to different $\alpha + N$ potentials 
(see Ref.~\cite{BT18} for details). 

\section{Comment on the exact-masses prescription}
To obtain non-vanishing $E1$ transitions in the two-body or potential model, 
experimental masses are used in the effective charge of $N = Z$ nuclei, 
\beq
Z_{\rm eff}^{({\rm E}1)} \propto \left(\frac{Z_1}{A_1} - \frac{Z_2}{A_2} \right)
\rightarrow
Z_{\rm eff}^{({\rm E}1)} \propto m_N \left(\frac{Z_1}{M_1} - \frac{Z_2}{M_2} \right)
\eeq
where $m_N$ is the nucleon mass and $Z_{1,2}$, $A_{1,2}$ and $M_{1,2}$ 
are the charges, mass numbers and experimental masses of the colliding nuclei, respectively. 
This exact-masses prescription is unfounded. \\
(i) $E1$ transitions would remain exactly forbidden in the $d(d,\gamma)^4$He reaction, 
in contradiction with {\it ab initio} calculations. \\
(ii) Using the mass expression $M = Am_N + (N-Z) \dem (m_n - m_p) - B(A,Z)/c^2$, 
effective charges would depend on the binding energies $B(A_{1,2},Z_{1,2})$, 
\beq
m_N \left(\frac{Z_1}{M_1} - \frac{Z_2}{M_2} \right) 
\approx \frac{1}{2m_N c^2} \left(\frac{B(A_1,Z_1)}{A_1} - \frac{B(A_2,Z_2)}{A_ 2} \right). 
\eeq
Binding energies per nucleon $B(A,Z)/A$ mostly depend on the main $T = 0$ components of the wave functions 
and not on the small $T = 1$ components physically responsible for the non vanishing of ``forbidden" $E1$ transitions. \\
(iii) $E1$ matrix elements would be unphysically sensitive to the long $T_f = 0$ $\alpha + d$ tail 
of the $^6$Li wave function. 
\section{Conclusion}
Isovector $E1$ transitions with $T = 1$ admixtures in the final state 
and $E2$ transitions explain the order of magnitude of the LUNA data \cite{ATM14,TAA17}, 
without any adjustable parameter. 
Isoscalar $E1$ transitions beyond the LWA are negligible for the $\alpha(d,\gamma)^6$Li reaction. 
The exact-masses prescription is not founded and should not be trusted for reactions between $N = Z$ nuclei, 
such as $\alpha(d,\gamma)^6$Li and $^{12}$C$(\alpha,\gamma)^{16}$O. 
A three-body model with $T = 1$ admixtures in both initial and final states should be developed. 
Microscopic six-body and {\it ab initio} calculations are difficult but possible 
and necessary for a deeper understanding of this reaction. 
\\
\\

\end{document}